\documentclass[aps,prc,twocolumn,groupedaddress,showpacs,10pt,dvipsnames]{revtex4-1}

\usepackage[utf8]{inputenc}
\usepackage{amsfonts}
\usepackage{amsmath}
\usepackage{amssymb}
\usepackage{bm}
\usepackage[pdftex]{graphicx}
\usepackage{xcolor}
\usepackage[normalem]{ulem}
\usepackage{comment}

\begin{document}

\newcommand{\tj}[6]{ \begin{pmatrix}
  #1 & #2 & #3 \\
  #4 & #5 & #6
 \end{pmatrix}}


\title{Identifying structures in the continuum: Application to $^{16}$Be} 




\author{J. Casal}
\email{casal@pd.infn.it}
\affiliation{European Centre for Theoretical Studies in Nuclear Physics and Related Areas (ECT$^*$), Villa Tambosi, Strada delle Tabarelle 286, I-38123 Trento, Italy}
\affiliation{Dipartimento di Fisica e Astronomia ``G.~Galilei'' and INFN - Sezione di Padova, Via Marzolo 8, I-35131 Padova, Italy}
\author{J. G\'omez-Camacho}
\affiliation{Departamento de F\'{\i}sica At\'omica, Molecular y Nuclear, Facultad de F\'{\i}sica, Universidad de Sevilla, Apartado 1065, E-41080 Sevilla, Spain} 
\affiliation{Centro Nacional de Aceleradores, U.~Sevilla, J.~Andalucía, CSIC, Tomas A.~Edison 7, E-41092 Sevilla, Spain} 


\date{\today}

\begin{abstract}
 \begin{description}
  \item[Background] The population and decay of two-nucleon resonances offer exciting new opportunities to explore dripline phenomena. A proper understanding of these systems requires a solid description of the three-body ($\text{core}+N+N$) continuum. The identification of a state with resonant character from the background of non-resonant continuum states in the same energy range poses a theoretical challenge.
  \item[Purpose] Establish a robust theoretical framework to identify and characterize three-body resonances in a discrete basis, and apply the method to the two-neutron unbound system $^{16}$Be.
  \item[Method] A resonance operator is proposed, which describes the sensitivity to changes in the potential. Resonances, understood as normalizable states describing localized continuum structures, are identified from the eigenstates of the resonance operator with large negative eigenvalues. For this purpose, the resonance operator is diagonalized in a basis of Hamiltonian pseudostates, which in the present work are built within the hyperspherical harmonics formalism using the analytical transformed harmonic oscillator basis. The energy and width of the resonance are determined from its time dependence. 
  \item[Results] 
   The method is applied to $^{16}$Be in a $^{14}\text{Be}+n+n$ model. An effective $\text{core}+n$ potential, fitted to the available experimental information on the binary subsystem $^{15}$Be, is employed.  The $0^+$ ground state resonance of $^{16}$Be presents a strong dineutron configuration. This favors the picture of a correlated two-neutron emission. Fitting the three body interaction to the experimental two-neutron separation energy $|S_{2n}|=1.35(10)$ MeV, the computed width is $\Gamma(0^+)=0.16$ MeV. From the same Hamiltonian, a $2^+$ resonance is also predicted with $\varepsilon_{r}(2^+)=2.42$ MeV and $\Gamma(2^+)=0.40$ MeV.
  \item[Conclusions] The dineutron configuration and the computed $0^+$ width are consistent with previous $R$-matrix calculations for the true three-body continuum.
  The extracted values of the resonance energy and width converge with the size of the pseudostate basis and are robust under changes in the basis parameters. This supports the reliability of the method in describing the properties of unbound $\text{core}+N+N$ systems in a discrete basis.
 \end{description} 
\end{abstract}


\maketitle


\section{Introduction}
Recent advances in radioactive ion beam physics allow us to explore dripline phenomena, where nuclear systems exhibit exotic properties~\cite{tanihata13} and unusual decay modes~\cite{pfutzner12}. Light nuclei away from stability are typically characterized by few or no bound states, with the continuum playing a fundamental role in shaping their structure properties~\cite{RomeroRedondo14} and reaction dynamics~\cite{Khalili03}. The coupling to the continuum is a key ingredient in theoretical models aiming to understand exotic nuclei~\cite{Austern87}, and the effects are especially crucial in the case of low-lying resonances~\cite{Cubero12,Kanungo15,JCasal15}. This has a strong imprint, for instance, on the electromagnetic response of two-neutron halo nuclei and other weakly bound systems~\cite{MRoGa05,JPFernandezGarcia13,JCasal16}. Moreover, resonant states in unbound nuclei can be populated in transfer or knockout reactions induced by exotic projectiles~\cite{spyrou12,Aksyutina13,Cavallaro17}. In this context, two-nucleon decays have attracted renewed attention~\cite{grigorenko09,spyrou12,oishi17,lovell17,JCasal18}. The description of few-body resonances, however, is not an easy task. 

The intuitive concept of a resonance corresponds to a range of continuum energy eigenstates that have a larger probability amplitude within the potential well, as compared to other non-resonant continuum states. This behavior in the continuum allows to construct a wave packet, as a combination of continuum states, that localizes the wave function inside the potential well, and cancels the oscillations outside~\cite{Danilin98}. This can be done efficiently, for a single-channel case, in the energy range within the vicinity of a phase shift that is a multiple of $\pi/2$. For a multichannel problem, such as three-body systems or two-body systems with core excitations, the exact scattering problem can be solved and resonances can be identified from the eigenphases obtained by diagonalizing the $S$-matrix~\cite{MRoGa09,Pinilla16,lovell17}. For three-body systems comprising several charged particles, the Coulomb problem requires very involved procedures~\cite{Nguyen12}. Recent \textit{ab initio} developments can also explore continuum structures and phase shifts, but the method demands 
large computational efforts and so far it is limited to relatively light systems~\cite{RomeroRedondo14,Calci16}. A possible alternative is to diagonalize the Hamiltonian in a square-integrable basis.

In general, the diagonalization of the few-body Hamiltonian in a discrete basis is referred to as pseudostate (PS) method~\cite{Tolstikhin97}, which provides a discrete set of positive-energy eigenstates representing the continuum. For this purpose, different bases can be employed~\cite{Desc03,Matsumoto03,MRoGa04,AMoro09,JCasal13}. As the basis size is increased, however, the density of pseudostates becomes larger, and the identification and study of resonances above the non-resonant background is difficult. It is possible to obtain phase shifts in a single-channel problem by using pseudostates and following the Hazi \& Taylor stabilization criteria~\cite{TaylorHazi76,JALay10}, but the extension to multichannel cases is not trivial. In Ref.~\cite{JCasal18}, it has been shown that three-body resonances, understood as localized continuum structures, can be associated to discrete eigenstates which are stable with respect to changes in the basis parameters. However, the method was restricted to a limited range of parameters which have to be determined by trial and error. Moreover, no information about the width of the state could be obtained from this representation of the continuum.

It is the purpose of this work to establish a more robust prescription to identify and characterize resonances using a discrete basis. A resonance operator will be introduced to single out localized continuum structures. Then, the resonance parameters $\varepsilon_r$ and $\Gamma$ will be determined from its time evolution. To apply the method to three-body systems, such as halo nuclei or two-nucleon emitters, the hyperspherical harmonic formalism~\cite{Zhukov93,Nielsen01} will be used. The method will be tested on the unbound $^{16}$Be nucleus, whose $0^+$ ground state has been recently claimed to decay via simultaneous two-neutron emission~\cite{spyrou12}. To assess the validity of the results, the resonance width will be compared to the exact three-body scattering calculations~\cite{lovell17} using the same interactions. In addition, predictions for the $1^-$ and $2^+$ continuum will be also presented.

The paper is structured as follows. In Sec.~\ref{sec:formalism}, the method to identify and characterize few-body resonances is introduced, together with the three-body framework used in this work. In Sec.~\ref{sec:application}, the formalism is applied to $^{16}$Be, and the reliability of the theoretical approach is discussed by comparing with previous results. Finally, Sec.~\ref{sec:conclusions} summarizes the main conclusions and outlines possible further applications.

\section{Theoretical framework}
\label{sec:formalism}

\subsection{Resonance operator}
\label{sec:operator}
Pseudostate (PS) approaches consist in solving a simple eigenvalue problem~\cite{Tolstikhin97}
\begin{align}
\widehat{H}|n\rangle &=\varepsilon_n|n\rangle,\label{eq:PSeq}\\
|n\rangle & = \sum_{i\beta}D_{i\beta}^n|i\beta\rangle,\label{eq:eigenH}
\end{align}
where $\{i,\beta\}$ label the radial excitation of the basis and the channel 
indexes (spins, orbital angular momenta and total angular momenta),  
respectively. The coefficients $D_{i\beta}^n$ are determined by diagonalizing the few-body Hamiltonian in a discrete basis~(e.g., \cite{HaziTaylor70,Desc03}), which requires just the kinetic energy and potential matrix elements
\begin{align}
T_{i\beta,i'\beta'}& =\langle i\beta|\widehat{T}|i'\beta'\rangle,\\
V_{i\beta,i'\beta'}&=\langle i\beta|\widehat{V}|i'\beta'\rangle.\label{eq:Vmatrix}
\end{align}
The solutions of Eq.~(\ref{eq:PSeq}) for negative-energy eigenvalues converge to the bound states of the system as the basis size is increased, while positive-energy eigenstates, or PS, provide a discrete representation of the continuum. Those that appear at relatively low energies can be extended non-resonant states occupying all the available configuration space covered by the basis functions, and thus being characterized by small values of the potential energy $V$ and the kinetic energy $T$. Alternatively, one can find localized PS exploring the range of the nuclear interaction, with large negative values of $V$ and comparable $T$, and typically associated with continuum structures such as resonances or virtual states. However, the diagonalization of $\widehat{H}$ in a large discrete basis mixes these two types of states~\cite{JCasal15}, which makes difficult the identification and study of continuum structures.

To address this problem, a procedure will be established to extract, from the large number of states that appear in the description of the continuum in a discrete basis, a non-stationary state which has properties that can be associated to a resonance. Thus, it will be required that:

1. The state representing the resonance should be specially sensitive to the interaction. Indeed, if there is no interaction, no resonances appear in the continuum.

2. The resonant state obtained should be robust versus changes in the basis set used. 

3. The resonant state should separate clearly from non-resonant continuum states.

4. The resonant state, in configuration space, should be a square-normalizable state, with a large probability to concentrate its components at short distances.

5. The energy distribution of the resonant state should be qualitative similar to a Breit-Wigner.

6. The time evolution of the state should resemble the exponential decay of a resonance.

Following criterion 1, it is possible to introduce the operator $\widehat{H}(\lambda) = \widehat{T} + \lambda \widehat{V}$, which, for $\lambda=1$, is simply reduced to the Hamiltonian. Then, assuming that localized continuum structures will be very sensitive to changes in the potential, the following operator is considered
\begin{equation}
\widehat{O}=\frac{d\widehat{H}(\lambda)}{d\lambda}\widehat{H}(\lambda)^{-1},
\end{equation}
i.e., the relative change of $\widehat{H}(\lambda)$ with respect to $\lambda$. However, the preceding operator is not Hermitian, as $\frac{d\widehat{H}(\lambda)}{d\lambda}$ does not commute with $\widehat{H}(\lambda)^{-1}$. Therefore, its symmetrized version evaluated at $\lambda=1$ is introduced,
\begin{equation}
\widehat{M}=\widehat{H}^{-1/2}\widehat{V}\widehat{H}^{-1/2}.
\label{eq:resop}
\end{equation}
The ansatz is that the eigenstates of the operator $\widehat{M}$ enable the identification of resonances. 
From the matrix elements of the potential given by Eq.~(\ref{eq:Vmatrix}), it is straightforward to write down the matrix elements of this new operator $\widehat{M}$ as
\begin{equation}
\langle n|\widehat{M}|n'\rangle={\varepsilon_n^{-1/2}}\langle n|\widehat{V}|n'\rangle {\varepsilon_{n'}^{-1/2}}.
\end{equation}
where $\langle n|\widehat{V}|n'\rangle$ can be easily computed from the expansion of the energy pseudostates in Eq.~(\ref{eq:eigenH}) as
\begin{equation}
\langle n|\widehat{V}|n'\rangle=\sum_{\beta\beta'}\sum_{ii'}D_{i\beta}^nD_{i'\beta'}^{n'}V_{i\beta,i'\beta'}.
\label{eq:potDD}
\end{equation}
The eigenstates of $\widehat{M}$ corresponding to the lowest eigenvalues $m$,
\begin{equation}
\widehat{M}|\psi\rangle=m|\psi\rangle,
\label{eq:Meq}
\end{equation}
are expected to characterize localized continuum structures such as resonances. According to criteria 2-4 above, this will be the case only if the state is clearly separated from the rest of the spectrum, it is concentrated at short distances and is stable under changes in the basis set used to describe the system. These states are expanded in eigenstates of the energy~(\ref{eq:eigenH}),
\begin{equation}
|\psi\rangle = \sum_{n}\mathcal{C}_{n}|n\rangle,\label{eq:eigenM}
\end{equation}
so that their energy distribution can be studied. This, together with its time dependence, will be used in the following sections to assess whether conditions 5-6 are fulfilled.

\begin{figure}
\centering
 \includegraphics[width=1\linewidth]{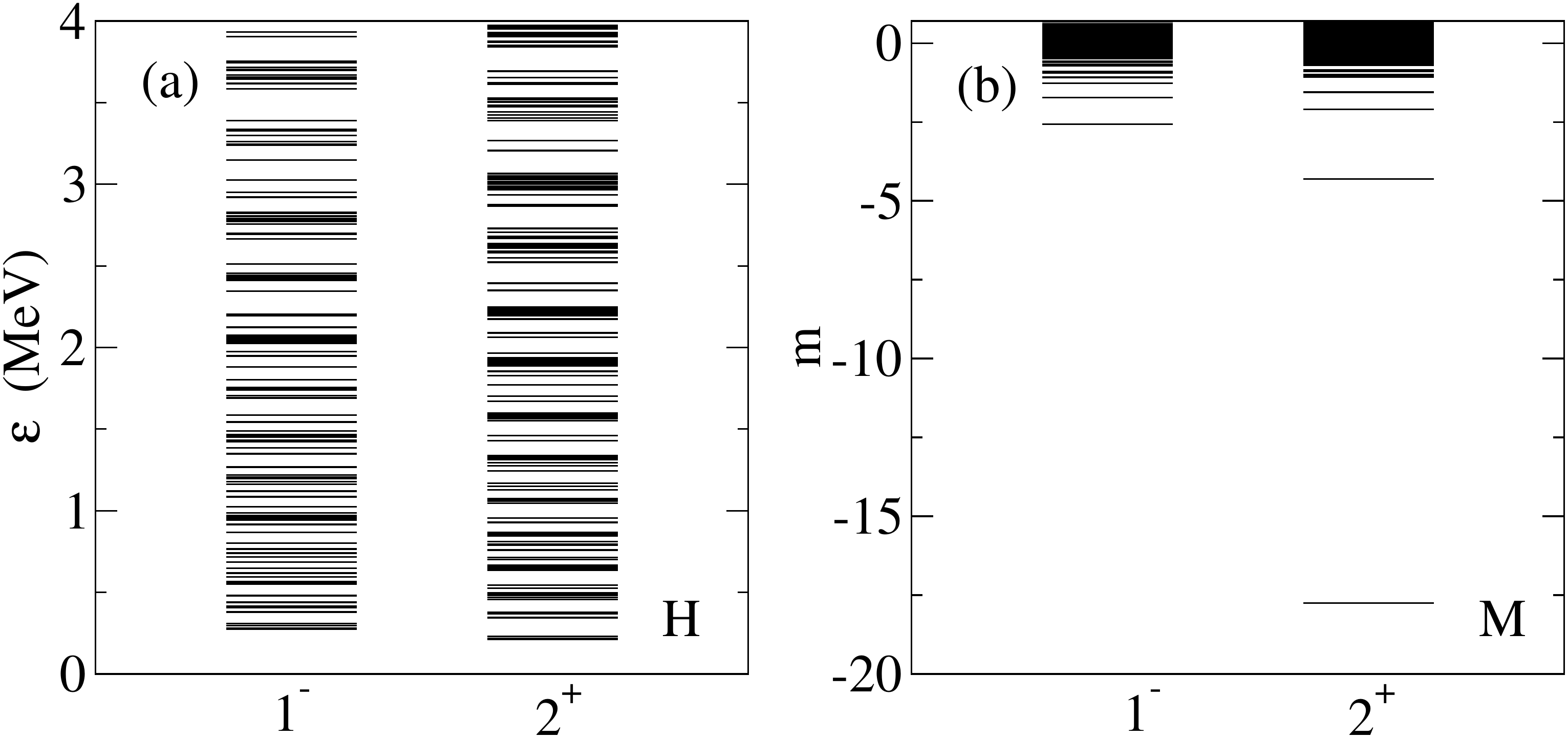}
 \caption{Typical spectra of (a) $\widehat{H}$ and  (b) $\widehat{M}$. This example corresponds to non-resonant $1^-$ states and a $2^+$ resonance.}
 \label{fig:example}
\end{figure}

In Fig.~\ref{fig:example}, the method is illustrated by studying the 1$^-$ and 2$^+$ states of the halo nucleus $^6$He~\cite{JCasal13}, in a large basis of pseudostates. The spectra of $\widehat{H}$ eigenvalues (left panel) is characterized in both cases by a large density of states, from which no resonant behavior can be disentangled. However the spectra of $\widehat{M}$ eigenvalues (right panel) shows that a 2$^+$ state is clearly separated from the rest, while this is not the case for 1$^-$ states. This 2$^+$ state corresponds to an eigenvalue of $m=-18$, indicating that the potential energy is significantly larger than the total energy. The result suggests that this state, which is not an eigenstate of the Hamiltonian, represents  a resonance. Besides, the fact that no 1$^-$ state can be similarly singled out shows that, within the model used to describe $^6$He~\cite{JCasal13}, there is no evidence of a 1$^-$ resonance.

\subsection{Time dependence and resonance parameters}
\label{sec:timedep}
As time evolves, states given by Eq.~(\ref{eq:eigenM}) become
\begin{equation}
|\psi(t)\rangle=\sum_n\mathcal{C}_ne^{-i\varepsilon_nt}|n\rangle,
\label{eq:tevol}
\end{equation}
where $t$ is the time divided by $\hbar$. This means that the initial state loses its character, and therefore it is possible to define a time-dependent amplitude from a given initial state $|\psi_o\rangle$ as
\begin{equation}
a(t)\equiv\langle\psi_o|\psi(t)\rangle=\sum_n|\mathcal{C}_n|^2e^{-i\varepsilon_nt}.
\label{eq:tamp}
\end{equation}
By definition, this amplitude equals 1 for $t=0$. For a resonance, one would expect~\cite{Satchler1}
\begin{equation}
a_r(t)=e^{-i\varepsilon_rt-\frac{\Gamma}{2} t},
\label{eq:tamp-res}
\end{equation}
given by the resonance energy $\varepsilon_r$ and its width $\Gamma$.
Note that the state $|\psi(t)\rangle$ is described as a combination of a finite number of energy eigenfunctions, and hence it cannot decay exponentially for long enough times. Nevertheless, for a physically motivated time range (e.g., associated to the time in which the resonance is produced in a reaction), one may require the time dependence of Eq.~(\ref{eq:tamp}) to be as close as possible to the resonance amplitude in Eq.~(\ref{eq:tamp-res}). Thus, the resonance parameters $\varepsilon_r$ and $\Gamma$ can be determined by minimizing the resonance quality parameter
\begin{equation}
\delta^2(\varepsilon_r,\Gamma)=\frac{\int_0^\infty W(t)\left|a(t)-a_r(t)\right|^2 dt}{\int_0^\infty W(t)\left|a(t)\right|^2 dt},
\label{eq:rqp}
\end{equation}
which has the meaning of a quadratic deviation. Here, $W(t)$ is a time profile describing the relevant time scale. For convenience, it can be parametrized simply as $W(t)=e^{-xt}$, where $x$ is a parameter with dimensions of energy. Thus, $\hbar/x$ corresponds to a relevant time scale for the resonance formation, such as a the collision time in which the resonance is produced. Note that small $x$ values will be related to long times associated to the decay of the resonance. In order to find the resonance parameters $\varepsilon_r$ and $\Gamma$ which best describe the time evolution of $a(t)$, Eq.~(\ref{eq:rqp}) can be minimized,
\begin{align}
\frac{\partial}{\partial \varepsilon_r}\delta^2(\varepsilon_r,\Gamma) & =0,\\
\frac{\partial}{\partial \Gamma}\delta^2(\varepsilon_r,\Gamma)& =0.
\end{align}
From these conditions one gets
\begin{equation}
\varepsilon_r=\sum_n\frac{\varepsilon_n\left|\mathcal{C}_n\right|^2}{Q_n(x)^2}\Big/\sum_n\frac{\left|\mathcal{C}_n\right|^2}{Q_n(x)^2},
\label{eq:er-eq}
\end{equation}
and
\begin{equation}
\frac{1}{\left(x+\Gamma\right)^2} = \sum_n\frac{\left|\mathcal{C}_n\right|^2 P_n(x)}{Q_n(x)^2},
\label{eq:gamma-eq}
\end{equation}
where
\begin{align}
Q_n(x)& =\left(x+\frac{\Gamma}{2}\right)^2+\left(\varepsilon_n-\varepsilon_r\right)^2,\label{eq:Qfact}\\
P_n(x)& =\left(x+\frac{\Gamma}{2}\right)^2-\left(\varepsilon_n-\varepsilon_r\right)^2.\label{eq:Pfact}
\end{align}
In practice, Eq.~(\ref{eq:er-eq}) can be solved iteratively to obtain $\varepsilon_r$ as a function of $\left(x+\Gamma/2\right)$. From this, and using Eq.~(\ref{eq:gamma-eq}), one gets $\left(x+\Gamma\right)$. In this way, the resonance parameters are obtained as a function of $x$, i.e., $\varepsilon_r(x)$ and $\Gamma(x)$.

Once the resonance parameters which best describe the time-dependent amplitude of the state have been determined, the quality of the resonance can be assessed from Eq.~(\ref{eq:rqp}). This, as a function of $x$, is given by
\begin{equation}
\delta^2(x)=\frac{F(x)-2E(x)+\left(x+\Gamma(x)\right)^{-1}}{F(x)},
\label{eq:rqp_x}
\end{equation}
where
\begin{equation}
F(x)=\sum_{nm}\left|\mathcal{C}_n\right|^2\left|\mathcal{C}_m\right|^2\frac{x}{x^2+\left(\varepsilon_n-\varepsilon_m\right)^2},
\end{equation}
\begin{equation}
E(x)=\sum_{n}\frac{\left(x+\Gamma(x)/2\right)\left|\mathcal{C}_n\right|^2}{Q_n(x)}.
\end{equation}
For large values of $x$, it is expected that $\delta^2(x)\rightarrow 0$, since the time profile $W(t)=e^{-xt}$ will explore very short times at which $a_r(t)$ and $a(t)$ trivially coincide. Small values of $x$, on the contrary, are more relevant to assess whether the state $|\psi(t)\rangle$ corresponds to a resonance, since they explore longer times.

\subsection{Three-body systems}

\begin{figure}[t]
\centering
 \includegraphics[width=0.28\linewidth]{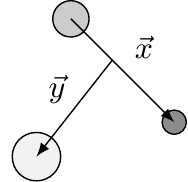}
 \caption{Jacobi coordinates for a three-body system.}
 \label{fig:Jacobi}
\end{figure}

The method will be applied to identify and characterize three-body resonances using the hyperspherical harmonics formalism~\cite{Zhukov93,Nielsen01}. The eigenstates of the three-body Hamiltonian are expanded as
\begin{align}
  |n\rangle\equiv\Psi_n(\rho,\Omega)& = \rho^{-5/2}\sum_{\beta}\left(\sum_i D_{i\beta}^{n} U_{i\beta}(\rho)\right)\mathcal{Y}_{\beta}(\Omega),\nonumber\\
  & = \rho^{-5/2}\sum_{\beta}R_\beta^n(\rho)\mathcal{Y}_\beta(\Omega)
 \label{eq:3bwf}
\end{align}
where $\rho^2=x^2+y^2$ is the hyperradius and $\Omega=\left\{\alpha,\widehat{x},\widehat{y}\right\}$ combines all the angular dependence, with $\tan\alpha=x/y$ the hyperangle. Here, $\left\{\boldsymbol{x},\boldsymbol{y}\right\}$ are the usual Jacobi coordinates in Fig.~\ref{fig:Jacobi}. Note that there are three possible choices of Jacobi coordinates, although a fixed set will be assumed here for simplicity. The index $i$ counts the number of basis functions, or hyperradial excitations, and the label $\beta\equiv\{K,l_x,l_y,l,S_x,j_{ab}\}j$ is typically referred to as channel, so that $R_\beta^n(\rho)$ is the radial wave function for each one. Functions $\mathcal{Y}_{\beta}(\Omega)$ are states of good total angular momentum $j$ following the coupling order
\begin{equation}
\mathcal{Y}_{\beta}(\Omega)=\left\{\left[\Upsilon_{Klm_l}^{l_xl_y}(\Omega)\otimes\kappa_{s_x}\right]_{j_{ab}}\otimes\phi_I\right\}_{j\mu}.
\label{eq:Upsilon}
\end{equation}
In this expression, $\boldsymbol{l}=\boldsymbol{l_x}+\boldsymbol{l_y}$, $S_x$ is the total spin of the two particles related by the $x$ coordinate, and $I$ represents the spin of the third particle, which is assumed to be fixed. The functions $\Upsilon_{Klm_l}^{l_xl_y}$ are the hyperspherical harmonics~\cite{Zhukov93}, and $K$ is the so-called hypermomentum. More details can be found, for instance, in Ref.~\cite{JCasal18}.

For the radial functions, the analytical transformed harmonic oscillator (THO)~\cite{JCasal13,JCasal14,JCasal16} basis is used. By performing a local scale transformation on the harmonic oscillator functions, 
\begin{equation}
  U_{i\beta}^{\text{THO}}(\rho)=\sqrt{\frac{ds}{d\rho}}U_{iK}^{\text{HO}}[s(\rho)].
\label{eq:Urho}
\end{equation}
the Gaussian asymptotic behavior is replaced by an exponential decay. Using the analytical form
\begin{equation}
s(\rho) = \frac{1}{\sqrt{2}b}\left[\frac{1}{\left(\frac{1}{\rho}\right)^{4} +
\left(\frac{1}{\gamma\sqrt{\rho}}\right)^4}\right]^{\frac{1}{4}},
\label{eq:LST}
\end{equation}
the parameters $b$ and $\gamma$ control the hyperradial extension of the basis, which is related to the density of pseudostates as a function of the energy. As shown in Refs.~\cite{JCasal13,JCasal18}, small $\gamma$ (or large $b$) values provide a higher concentration of discretized continuum states close to the breakup threshold. The level density after diagonalization is controlled by the ratio $\gamma/b$~\cite{JALay10,JCasal13}, therefore it is reasonable to fix one ($b$) and use the other as a parameter ($\gamma$), as in Ref.~\cite{JCasal18}.

The energy pseudostates $|n\rangle=\Psi_n$ are obtained by diagonalizing the three-body Hamiltonian in a given THO basis. This requires the hyperradial coupling potentials
\begin{equation}
V_{\beta'\beta}(\rho)=\left\langle \mathcal{Y}_{\beta }(\Omega)\Big|V_{12}+V_{13}+V_{23} \Big|\mathcal{Y}_{\beta'}(\Omega) \right\rangle + \delta_{\beta\beta'}V_{\rm 3b}(\rho),
\label{eq:vcoupl}
\end{equation}
where $V_{ij}$ are the corresponding two-body interactions, fitted by the known experimental information on the binary subsystems, and $V_{\rm 3b}(\rho)$ is a phenomenological three-body force. The latter is typically introduced to account for effects not explicitly included in a strict three-body picture~\cite{IJThompson04,MRoGa05,RdDiego10,JCasal15}, and its parameters can be fixed to shift the three-body energies without a significant change in the structure of the states. From the hyperradial couplings, the potential matrix elements required in Eq.~(\ref{eq:potDD}) are simply
\begin{equation}
V_{i\beta,i'\beta'}=\int d\rho U_{i\beta}^{\text{THO}}(\rho) V_{\beta,\beta'}(\rho) U_{i'\beta'}^{\text{THO}}(\rho).
\end{equation}
Note that the expansion~(\ref{eq:3bwf}) involves infinite sums over $\beta$ and $i$. However, calculations are typically truncated by fixing a maximum hypermomentum $K_{max}$ and a maximum number of  hyperradial excitations $i_{max}$ in each channel. These parameters should be chosen large enough to provide converged results. Note that fixing $K$ restricts $l_x$ and $l_x$ values to $l_x+l_y \le K$~\cite{Zhukov93}, such that no additional truncation is needed. 

\section{Application to $\boldsymbol{^{16}}$Be}
\label{sec:application}

\begin{figure}[t]
\centering
 \includegraphics[width=0.8\linewidth]{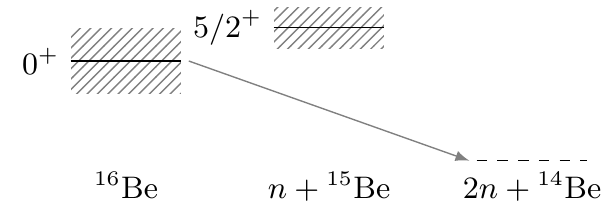}
 \caption{Decay of $^{16}$Be. Relative energies from Refs.~\cite{spyrou12,snyder13}.}
 \label{fig:16bedecay}
\end{figure}

The $^{16}$Be system has been recently populated in the $^{17}\text{B}(p,2p){^{16}\text{Be}}$ reaction at MSU~\cite{spyrou12}. Its ground state has been identified as a broad $0^+$ resonance characterized by $S_{2n}=-1.35(10)$ MeV and $\Gamma=0.8(1)$ MeV. The energy and angular correlations of the emitted neutrons suggested a simultaneous decay. This was consistent with the available information on the binary subsystem $^{15}$Be~\cite{snyder13}. In Fig.~\ref{fig:16bedecay}, the decay path is illustrated. The structure of the $^{16}$Be ground state was studied using the hyperspherical $R$-matrix method~\cite{lovell17}, showing a dominant dineutron configuration in the resonance wave function. This result was confirmed recently from a simpler pseudostate approach~\cite{JCasal18}. 

\subsection{Three-body model $\boldsymbol{^{14}\text{Be}+n+n}$}
\label{sec:3bmodel}

In order to test the suitability of the method, the three-body $^{14}\text{Be} + n + n $ problem is solved in this work using the same interactions as in Refs.~\cite{lovell17,JCasal18}. For the $n$-$n$ interaction, the Gogny-Pires-Tourreil (GPT)~\cite{GPT} potential is used. This parametrization includes central, spin-orbit and tensor terms, and reproduces nucleon-nucleon scattering observables up to 300 MeV. For the $^{14}\text{Be}+n$ interaction, an $l$-dependent Woods-Saxon potential with central and spin-orbit terms was adjusted to reproduce the available experimental information on $^{15}$Be, i.e., a $d_{5/2}$ ground state 1.8(1) MeV above the neutron-emission threshold and a width of 0.58(20) MeV~\cite{snyder13}. Note that these numbers, as well as the ground-state energy of $^{16}$Be, might change once new data with better energy resolution is available~\cite{Marques18}. The $^{14}\text{Be}+n$ potential parameters are given in Ref.~\cite{JCasal18}. This potential produces $1s_{1/2}$, $1p_{3/2}$ and $1p_{1/2}$ Pauli states that have to be projected out for the three-body calculations. For this purpose, different prescriptions can be adopted~\cite{IJThompson00}. Here, as in Refs.~\cite{lovell17,JCasal18}, a supersymmetric potential is constructed~\cite{Baye87}. This leads to a phase-equivalent potential with shallow $s$ and $p$ terms, which do not support $1s$ and $1p$ bound states. Although this inert-core approximation might not be the most realistic picture to describe $^{15,16}$Be, it is important in this context to use the same prescription when dealing with Pauli states, thus ensuring a sensible comparison between different three-body calculations. Exploring the effect of different Pauli treatments is beyond the scope of the present work. In addition to the binary interactions, the phenomenological three-body force introduced by Eq.~(\ref{eq:vcoupl}) is also included. This is Gaussian potential with $\rho_{\rm 3b}=6$ fm and $v_{\rm 3b}=-0.9$ MeV, and it has been adjusted to reproduce the two-neutron separation energy in $^{16}$Be once calculations are converged.

\begin{figure}
\centering
 \includegraphics[width=0.9\linewidth]{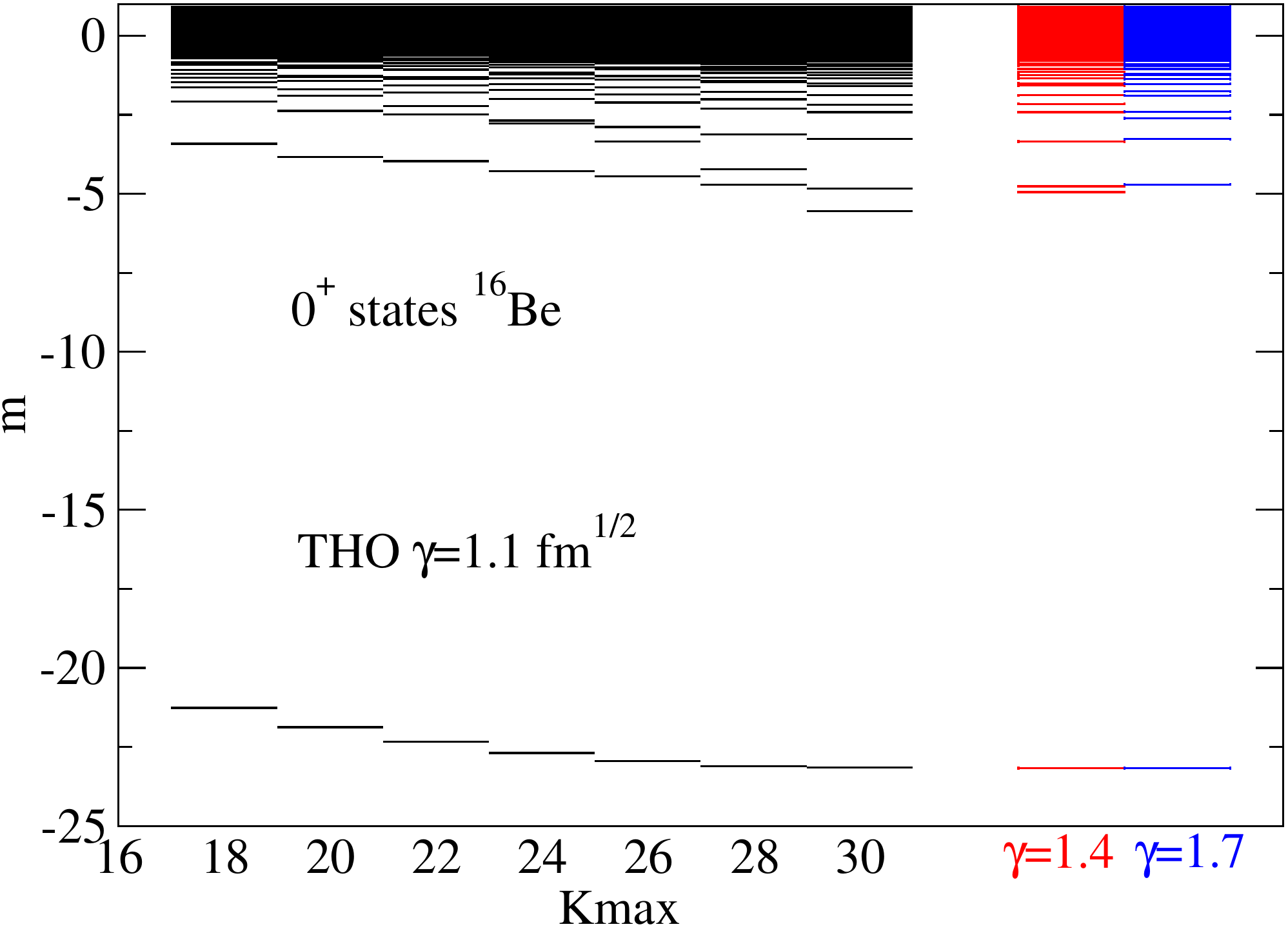}
 \caption{(Color online) Eigenvalues $m$ for $0^+$ states as a function of $K_{max}$. Calculations correspond to $b=1.4$ fm, $\gamma=1.1$ fm$^{1/2}$ and $i_{max}=20$. In the last two spectra, the converged results with $\gamma=1.4$ and 1.7 fm$^{1/2}$ are also shown.}
 \label{fig:eigenvaluesM}
\end{figure}

Calculations are performed in two steps. First, the three-body Hamiltonian in the Jacobi-T set, where the two valence neutrons are related by the $x$ coordinate, is diagonalized for $0^+$ states using a set of THO function with fixed basis parameters. Then, the resulting energy eigenstates are used to diagonalize the resonance operator $\widehat{M}$ in Eq.~(\ref{eq:resop}). In Fig.~\ref{fig:eigenvaluesM}, the $0^+$ eigenvalues $m$ are shown as a function of the maximum hypermomentum $K_{max}$ defining the model space. These calculations were performed with THO basis parameters $b=1.4$ fm and $\gamma=1.1$ fm$^{1/2}$, and $i_{max}=20$ hyperradial excitations, which were enough to achieve convergence. From this plot, it is clear that the operator $\widehat{M}$ divides the spectrum into two regions: a) the upper part, associated to spread, non-resonant continuum states, and b) the lowest, localized eigenstate which might be interpreted as a resonance. The latter shows a fast convergence, which resembles that of three-body bound states (e.g.~Refs.~\cite{JCasal13,JCasal14}). In Fig.~\ref{fig:eigenvaluesM}, calculations with different values of the $\gamma$ parameter (and $K_{max}=30$) are also presented. It is shown that the lowest eigenstate of $\widehat{M}$ is very stable with respect to changes in the basis parameters, provided the number of basis functions is large enough. This provides a robust representation of the resonance, independent of the basis extension, and it is a clear difference with respect to the calculations in Ref.~\cite{JCasal18}.

\begin{figure}
\centering
 \includegraphics[width=0.9\linewidth]{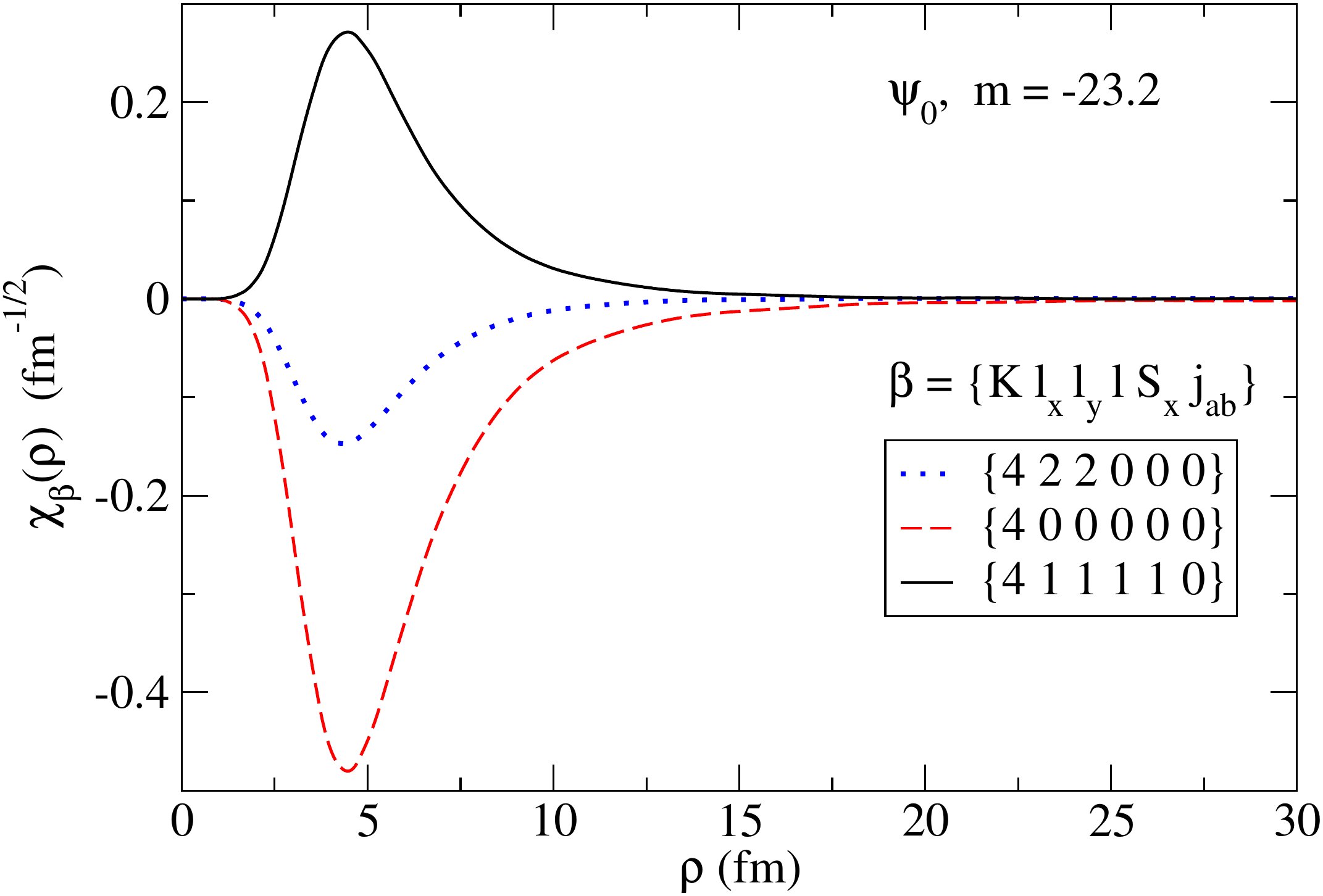}
 \caption{(Color online) Radial dependence of the channel wave functions for the lowest $0^+$ eigenstate of $\widehat{M}$. The three most important channels are shown.}
 \label{fig:wf}
\end{figure}

\begin{figure}
\centering
 \includegraphics[width=0.87\linewidth]{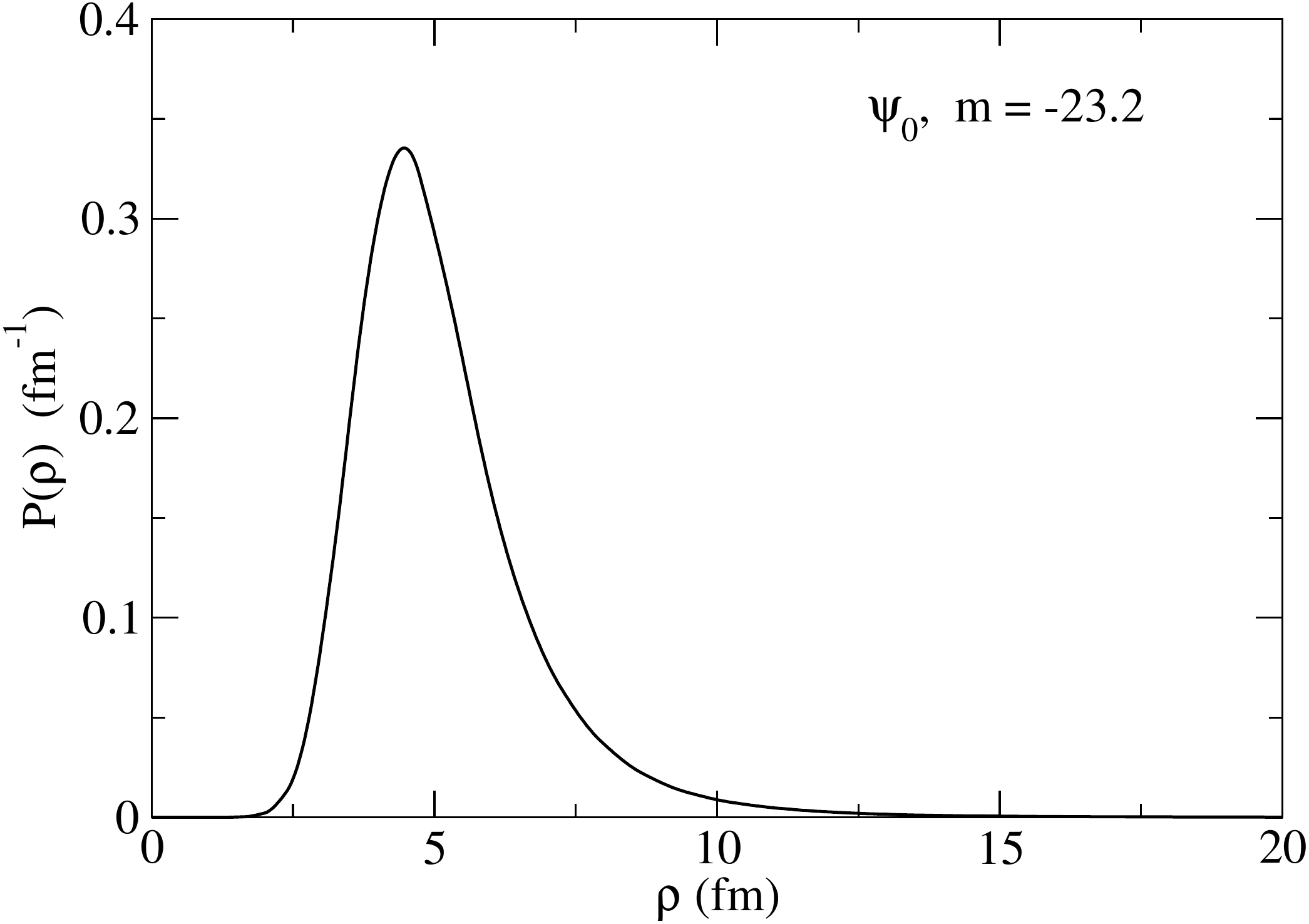}
 \caption{Hyperradial wave-function probability for the lowest $0^+$ eigenstate of $\widehat{M}$.}
 \label{fig:probrho}
\end{figure}

\begin{figure}
\centering
 \includegraphics[width=0.8\linewidth]{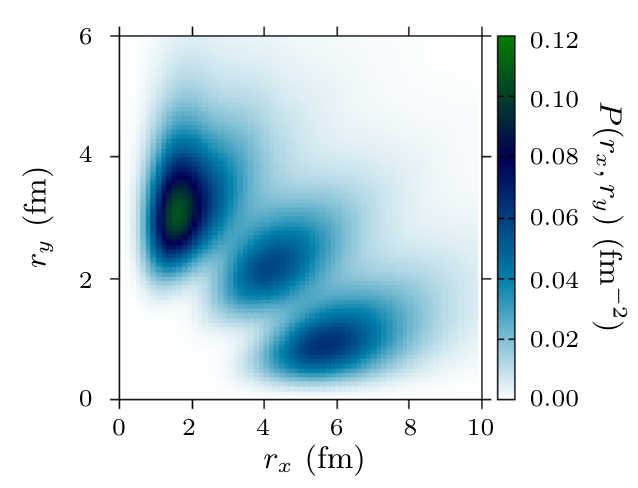}
 \caption{(Color online) Probability of the lowest $0^+$ state, with the scale in fm$^{-2}$, as a function of $r_x\equiv r_{nn}$ and $r_y\equiv r_{(nn)\text{-}{^{14}\rm Be}}$.}
 \label{fig:xyprob}
\end{figure}

By combining Eqs.~(\ref{eq:eigenM}) and~(\ref{eq:3bwf}), the lowest eigenstate of $\widehat{M}$ can be written as
\begin{equation}
 \psi_0(\rho,\Omega)=\rho^{-5/2}\sum_{\beta} \chi_{\beta 0}(\rho)\mathcal{Y}_{\beta}(\Omega),
 \label{eq:reswf}
\end{equation}
where $\chi_\beta(\rho)$ are the radial wave functions obtained after adding up the contributions from different energy eigenstates,
\begin{equation}
 \chi_{\beta}(\rho)=\sum_{n} \mathcal{C}_{n} R_{\beta}^n(\rho)=\sum_{n} \mathcal{C}_{n} \sum_i D_{i\beta}^{n} U_{i\beta}^{\rm THO}(\rho).
 \label{eq:channelwf}
\end{equation}
The most important channels in this expansion are shown in Fig.~(\ref{fig:wf}), where the dominance of an $l_x=0$ component in the Jacobi-T system can be noticed. In Fig.~(\ref{fig:probrho}), the total wave-function probability is presented. The state is localized at short hiperradi, a behavior expected for a resonance described in a discrete basis. The spatial correlations between the valence neutrons are shown in Fig.~\ref{fig:xyprob}, which presents three local maxima. The dominant one corresponds to the so-called dineutron configuration, with the two neutrons close to each other at some distance from the core. This result is consistent with those presented in Refs.~\cite{lovell17,JCasal18} for the ground-state resonance of $^{16}$Be. The second maximum in Fig.~\ref{fig:xyprob} corresponds to the cigar-like configuration, also observed in two-neutron halo systems~\cite{Zhukov93}, with the two neutrons far from each other but close to the core. Last, a third peak appears between the dineutron and cigar-like components, where the three particles are more equally spaced. This structure is sometimes called triangle configuration~\cite{lovell17}.

\subsection{Time evolution and resonance parameters}
\label{sec:timeevol}

\begin{figure}
\centering
 \includegraphics[width=0.97\linewidth]{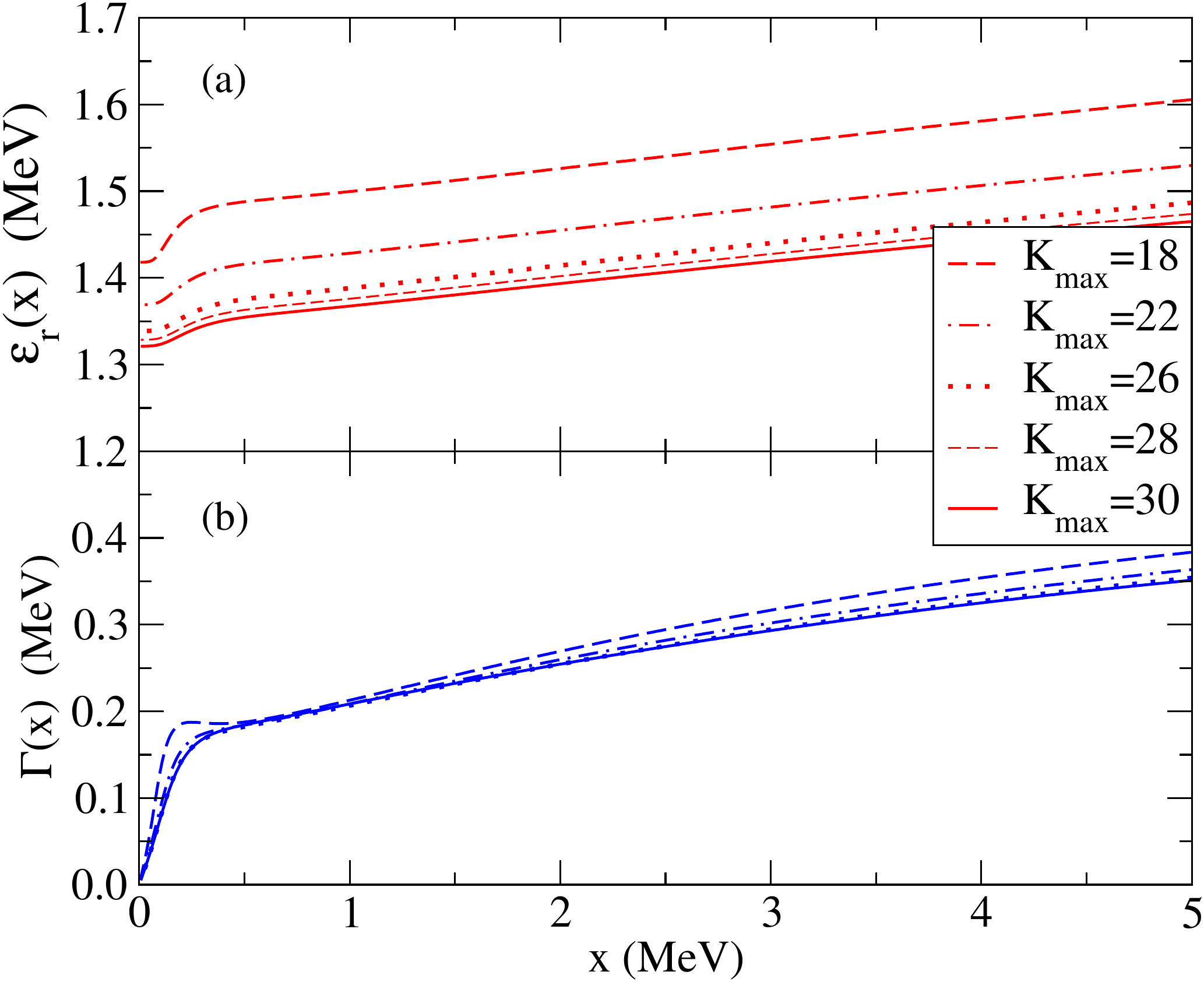}
 \caption{(Color online) Resonance parameters (a) $\varepsilon_r(x)$ and (b) $\Gamma(x)$. Convergence with respect to the maximum hypermomentum. With $K_{max}=30$, the resonance energy is converged within a 1\% difference.}
 \label{fig:res_kmax}
\end{figure}

\begin{figure}
\centering
 \includegraphics[width=0.97\linewidth]{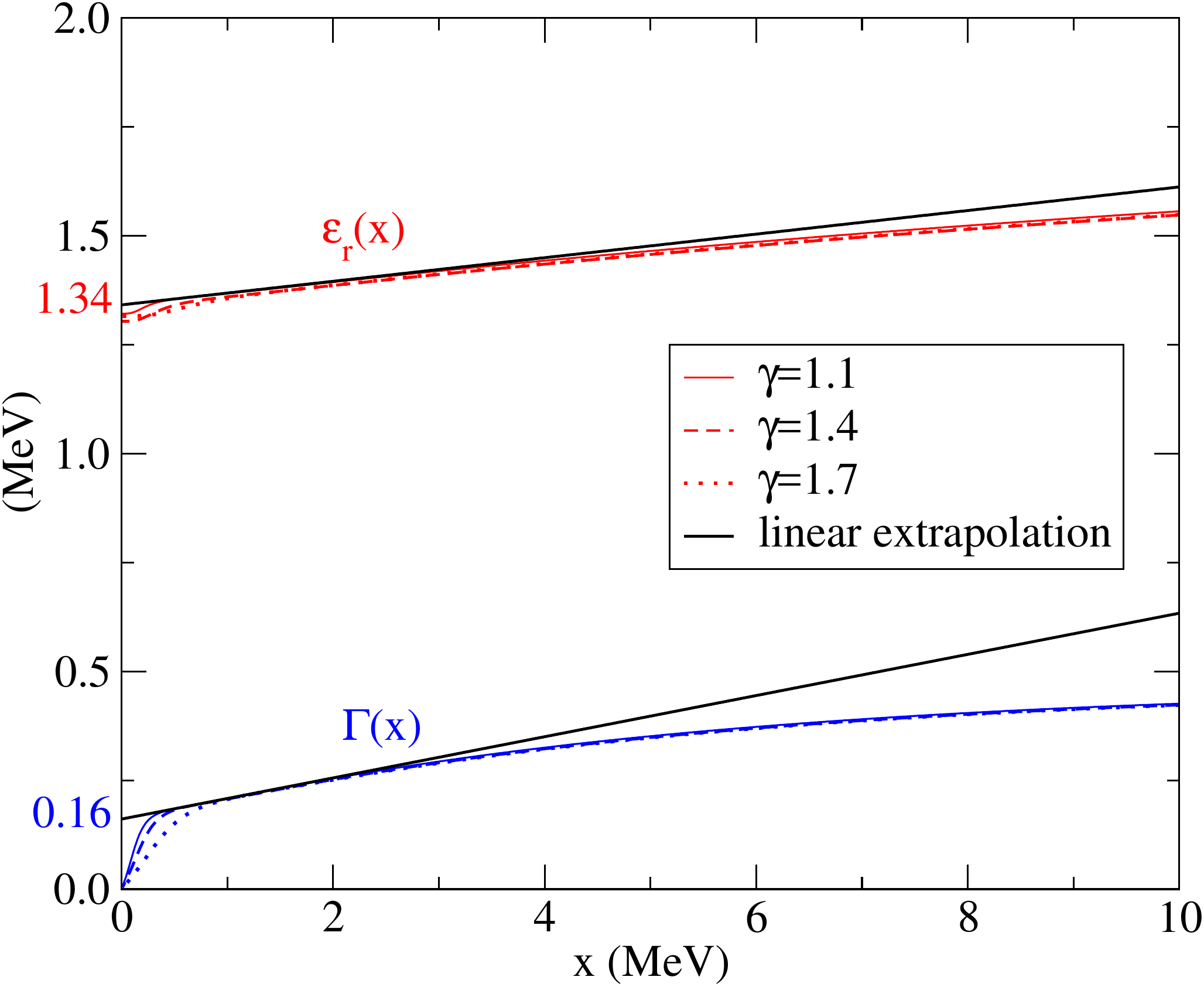}
 \caption{(Color online) Resonance parameters $\varepsilon_r(x)$ and $\Gamma(x)$ for three different THO bases, defined by their $\gamma$ parameter, together with a linear extrapolation for small values of $x$.}
 \label{fig:extrapolate}
\end{figure}

In the preceding section, the lowest eigenstate of the operator $\widehat{M}$ corresponding to $0^+$ states in $^{16}$Be has been identified as the ground-state resonance of this system. Its resonance parameters, i.e., $\varepsilon_r$ and $\Gamma$, are yet to be determined. This can be achieved by following the formalism presented in Sec.~\ref{sec:timedep}. The procedure yields the energy and width as a function of the parameter $x$, whose small values can be associated to long decay times. The energy and width functions so obtained present also a rather fast convergence pattern with respect to $K_{max}$, as can be seen in Fig.~\ref{fig:res_kmax}. These are obtained by solving Eqs.~(\ref{eq:er-eq}) and~(\ref{eq:gamma-eq}) iteratively. The energy functions for $K_{max}=28$ and 30 differ by less than $1\%$. The small effect from higher $K$ values can be effectively simulated by fitting the $K_{max}=30$ result to the experimental two-neutron separation energy in $^{16}$Be. Note that the three-body force employed in the preceding section was fitted so that the resonance energy in this plot is close to the experimental two-neutron separation energy in $^{16}$Be. With the adopted $K_{max}$ value, the width function is fully converged. In Fig.~\ref{fig:res_kmax}, it is shown that the energy and width functions follow approximately a linear trend, characterized by a small slope, for small values of $x$. Then, a sudden drop of the resonance width is observed for $x$ values close to zero. This decrease can be easily explained from Eqs.~(\ref{eq:gamma-eq}) and~(\ref{eq:Qfact}). The limit $\Gamma(0)\longrightarrow 0$ occurs when a discrete energy $\varepsilon_n$, with a non-vanishing amplitude $|\mathcal{C}_n|^2$, matches the resonance energy $\varepsilon_r$. Due to the discrete nature of the basis, this will likely occur if the median of the energy distribution characterizing the state is precisely an eigenvalue of the Hamiltonian. This issue can be overcome by increasing the level density around the resonance, i.e., by changing the THO parameters controlling the radial extension of the basis. To illustrate this, the energy and width functions are shown in Fig.~\ref{fig:extrapolate} for three different values of the transformation parameter $\gamma$. It is observed that, for smaller $\gamma$ (i.e., larger level densities at low energies) the linear trend explores smaller values of $x$ before the sudden decrease in the width. Therefore, this behavior can be extrapolated, providing an upper limit for both the resonance energy and the width. This is also shown in Fig.~\ref{fig:extrapolate}. Following this prescription, the parameters describing the resonance, at $x=0$, are $\varepsilon_r(0^+)=1.34$ MeV and $\Gamma(0^+)=0.16$ MeV. The latter is very close to the width obtained in Ref.~\cite{lovell17} using the hyperspherical $R$-matrix method to solve the true three-body scattering problem, 0.17 MeV. This confirms that the method here presented to identify and characterize resonances in a discrete basis, using the definition of the resonance operator $\widehat{M}$ and the time evolution of its lowest eigenstate, provides reasonable results.

Note that the resonance width obtained both in the present work and in the previous $R$-matrix calculations are significantly smaller than the reported experimental value of 0.8 MeV~\cite{spyrou12}. In Ref.~\cite{lovell17}, this difference was attributed to the experimental resolution. However, there might be deficiencies in the three-body model affecting the resonance width, such as the inert-core approximation or the treatment of Pauli states introduced in Sec.~\ref{sec:3bmodel}. There is also the possibility that the experimental energy distribution contained the contribution from two unresolved resonances, e.g.~the $0^+$ ground state and the first $2^+$ excited state~\cite{Marques18}. New experimental data with improved energy resolution could help in this context.

\begin{figure}
\centering
 \includegraphics[width=0.95\linewidth]{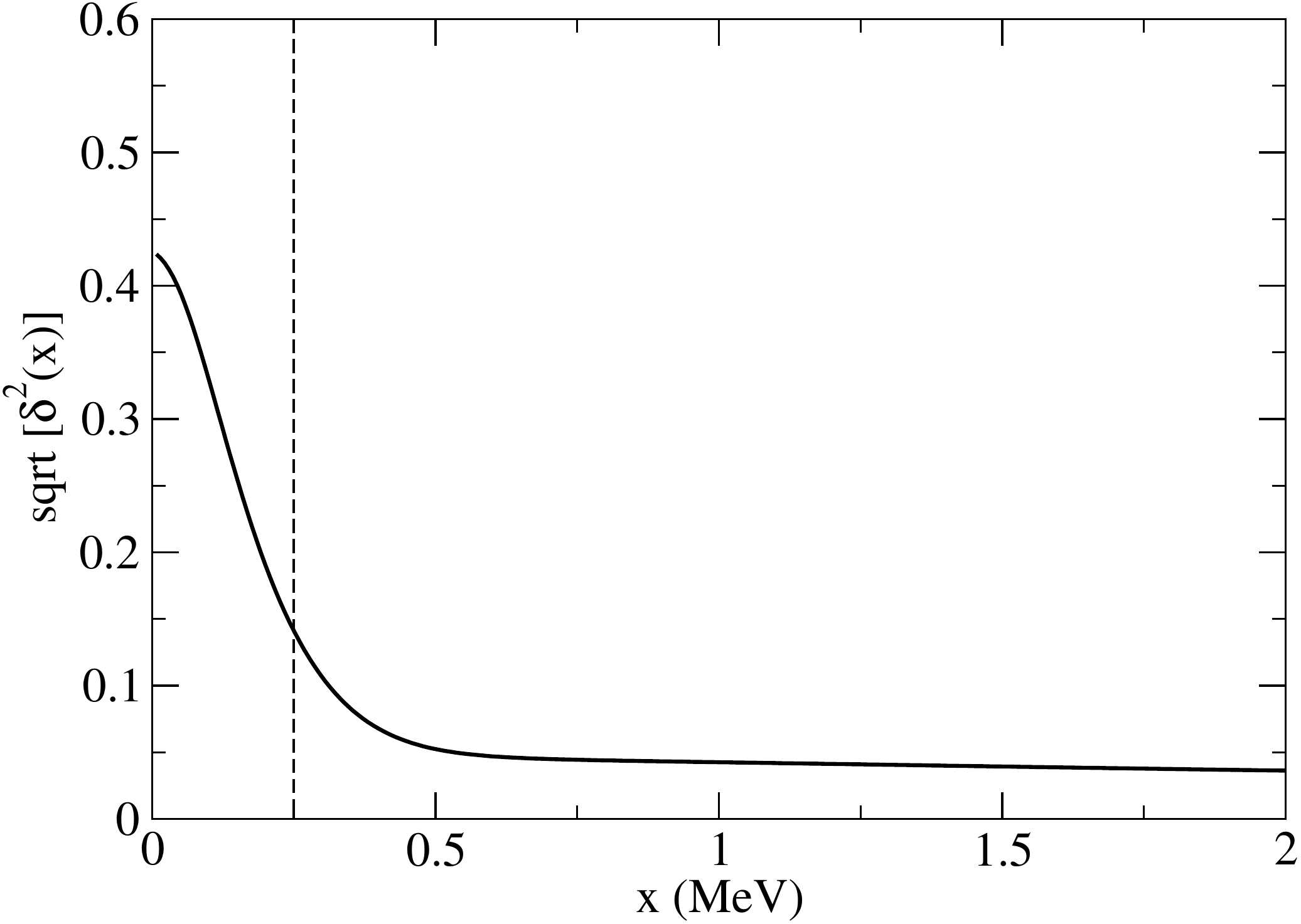}
 \caption{Square root of the quality parameter for the $0^+$ resonance in $^{16}$Be. The value $x_0=0.25$ MeV at which the resonance width $\Gamma(x)$ deviates from a linear character is highlighted.}
 \label{fig:quality}
\end{figure}

\begin{figure}
\centering
 \includegraphics[width=0.95\linewidth]{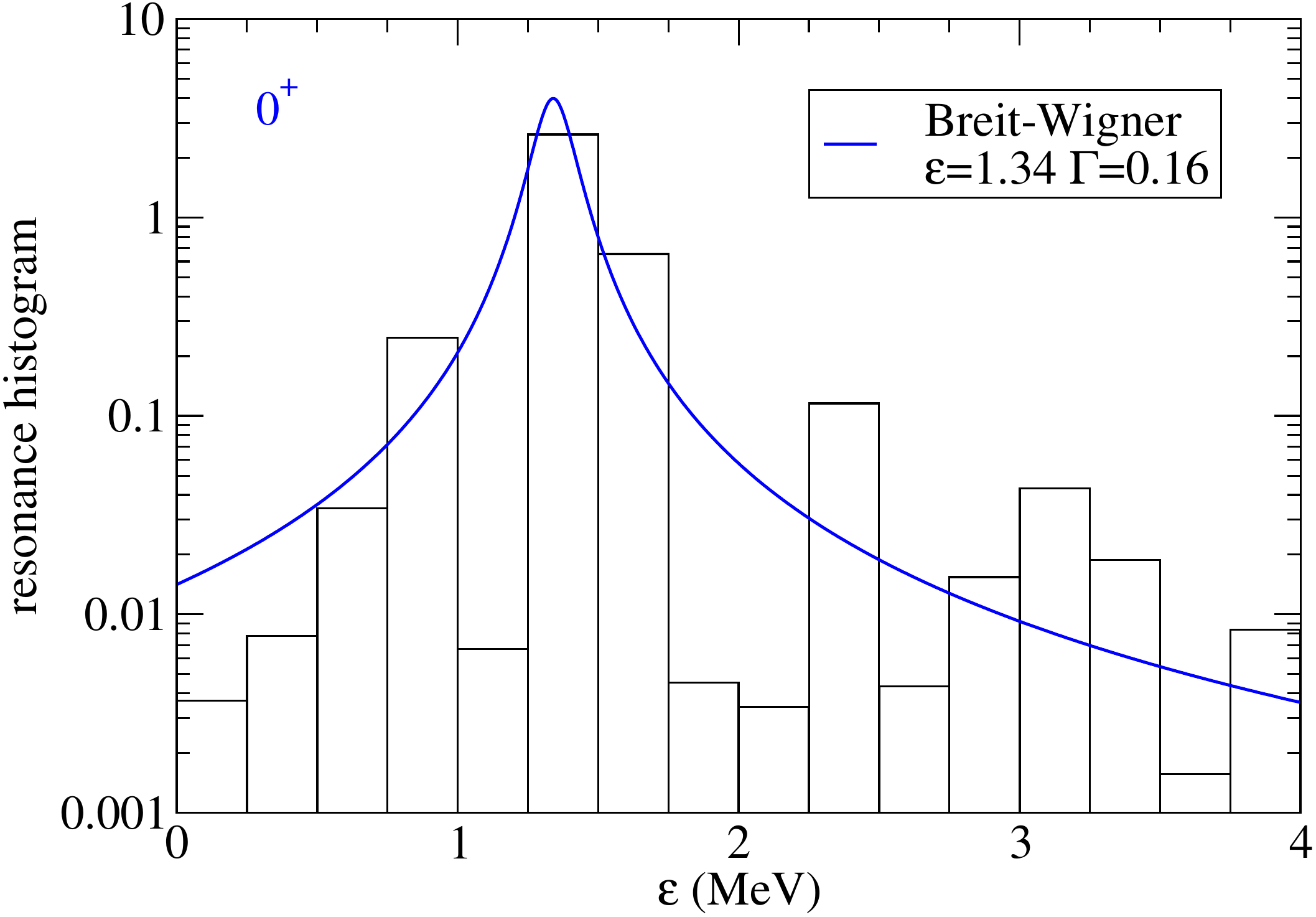}
 \caption{Energy distribution of the $0^+$ ground-state resonance, built as a histogram with a 0.25 MeV step. The solid line is a Breit-Wigner shape with the adopted resonance parameters $\varepsilon_r(0^+)=1.34$ MeV and $\Gamma(0^+)=0.16$ MeV.}
 \label{fig:histo0}
\end{figure}

Three-body resonances are multichannel states which do not necessary follow a typical Breit-Wigner shape. The resonance quality parameter introduced in Eq.~(\ref{eq:rqp}) is a measure of the quadratic deviation from this behavior. By using Eq.~(\ref{eq:rqp_x}), this quantity can be easily computed from the energy and width functions $\varepsilon_r(x)$ and $\Gamma(x)$. This is shown in Fig.~\ref{fig:quality}. As discussed in Sec.~\ref{sec:timedep}, the quality parameter at large $x$ is trivially zero, since the exponential $W(x)=e^{-xt}$ explores very short times. The limit $x\longrightarrow 0$, however, is not interesting either due to the sudden drop in the resonance parameters produced by the discrete nature of the basis. To get an idea about the quality of the resonance, one can look at the value of $\delta^2(x)$ where $\Gamma(x)$ deviates from the linear trend, i.e., $x_0=0.25$ MeV in this case. This gives a small value of $\displaystyle\sqrt{\delta^2(x_0)}=0.14$, which means that the deviation from a single-channel resonance is of the order of 14\%.
A histogram with the energy distribution of the resonance, corresponding to Eq.~(\ref{eq:eigenM}), is shown in Fig.~\ref{fig:histo0} using $x_0$ as the step. The distribution is slightly asymmetric, but shows a trend that can be qualitatively reproduced by a Breit-Winger resonance with parameters $\varepsilon_r(0^+)=1.34$ MeV and $\Gamma(0^+)=0.16$ MeV. This is consistent with the reported small value of $\delta^2(x_0)$.

\subsection{Prediction of resonances: $\boldsymbol{1^-}$ and $\boldsymbol{2^+}$ states.}
\label{sec:2plus}

\begin{figure*}
\centering
 \includegraphics[width=0.7\linewidth]{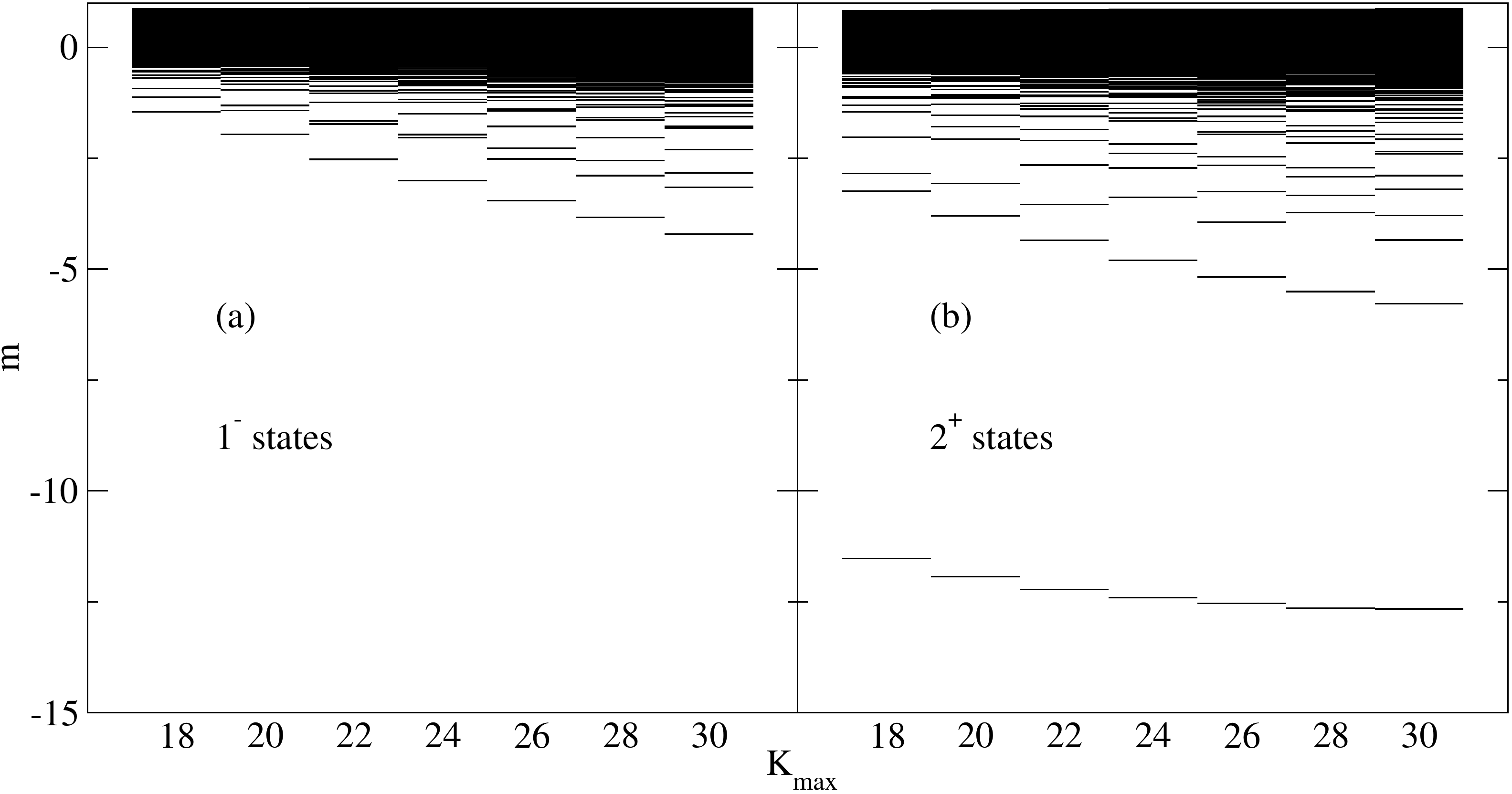}
 \caption{Eigenvalues $m$ for (a) $1^-$ and (b) $2^+$ states as a function of $K_{max}$.}
 \label{fig:1m2p}
\end{figure*}

\begin{figure}
\centering
 \includegraphics[width=0.97\linewidth]{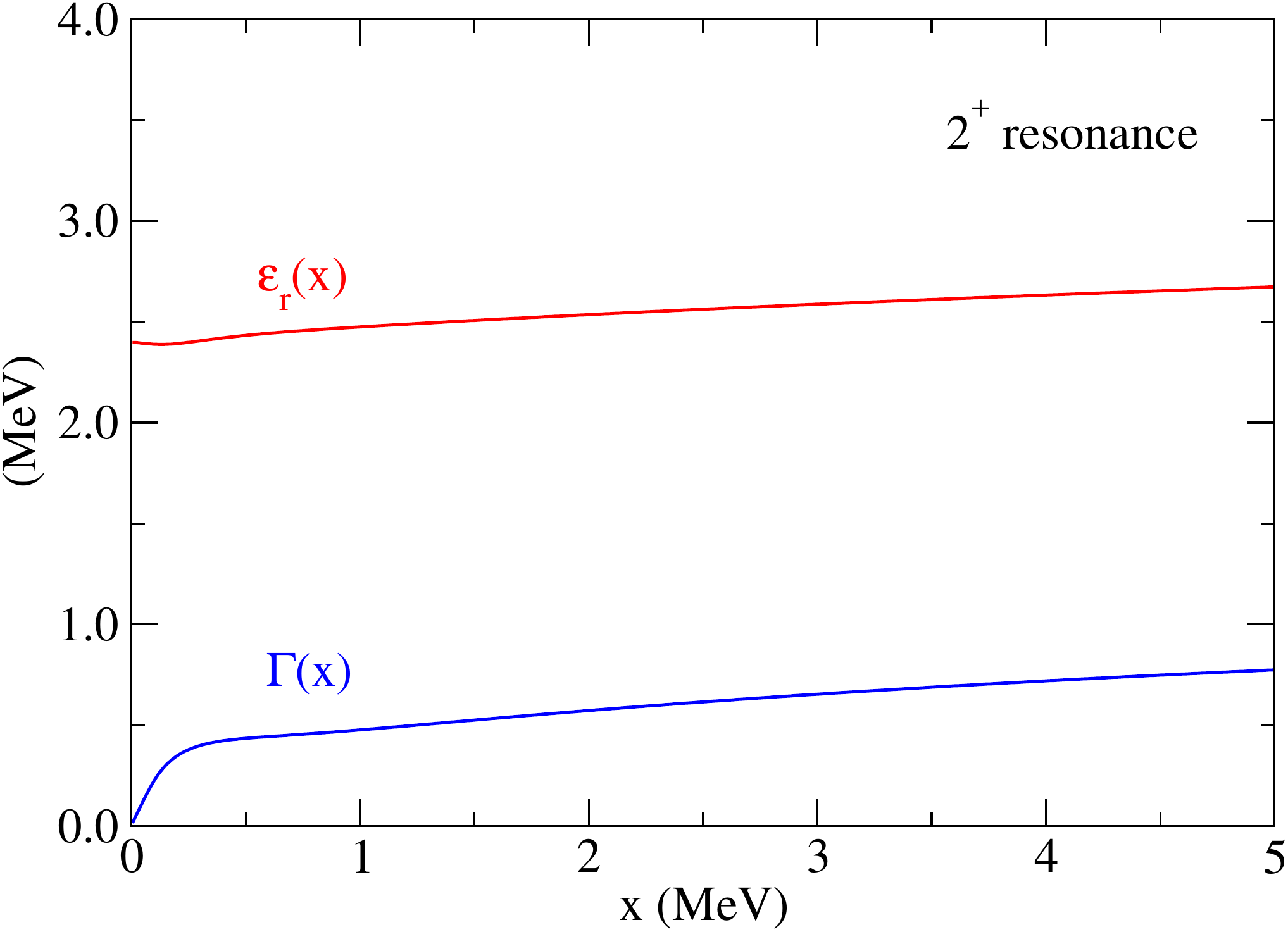}
 \caption{(Color online) Energy and width functions for the $2^+$ resonance in $^{16}$Be.}
 \label{fig:2plus}
\end{figure}

\begin{figure}
\centering
 \includegraphics[width=0.95\linewidth]{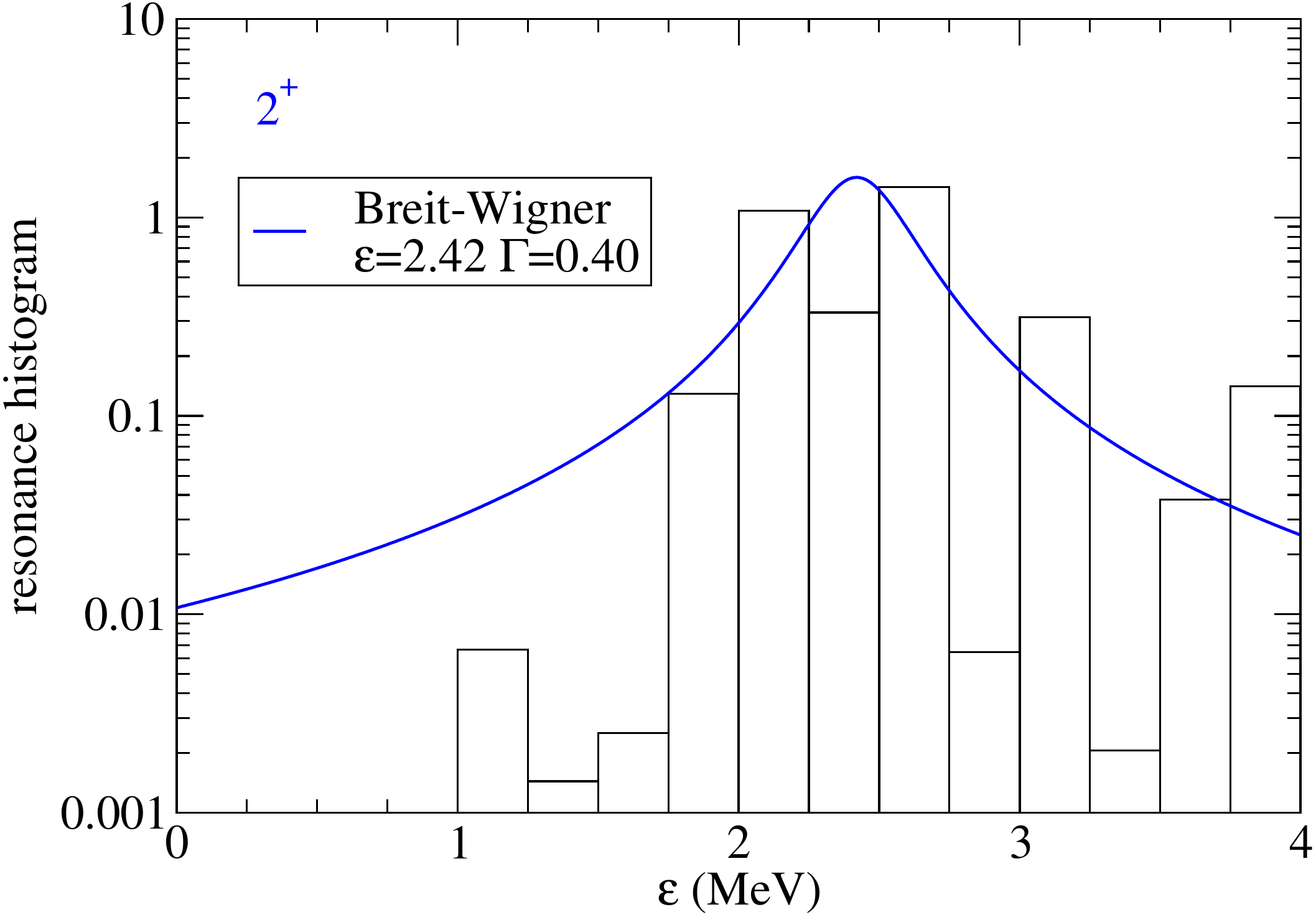}
 \caption{(Color online) The same as Fig.~\ref{fig:histo0} but for the $2^+$ resonance.}
 \label{fig:histo2}
\end{figure}

The present formalism has been applied to characterize the $0^+$ ground-state resonance of $^{16}$Be, but the method can be easily extended to study the three-body continuum for different angular momentum states. The eigenvalues of $\widehat{M}$ for $1^-$ and $2^+$ states, from the same Hamiltonian and using the same THO basis and model space, are shown in Fig.~\ref{fig:1m2p} as a function of $K_{max}$. 
The procedure allows to identify a clear $2^+$ resonance in the $m$ eigenvalue plot, as opposed to what it is observed for the $1^-$ case. 
The energy and width functions for this state are shown in Fig.~\ref{fig:2plus}, 
from which the linear extrapolation gives $\varepsilon_r(2^+)=2.42$ MeV and $\Gamma(2^+)=0.40$ MeV. The quality of the resonance in this case is $\displaystyle\sqrt{\delta^2(x_0)}=0.25$, so a significant deviation is observed, as compared to that of the $0^+$ ground state. 
This is more clear in the energy distribution of the state given in Fig.~\ref{fig:histo2}, compared to the Breit-Wigner shape with the adopted resonance parameters.

It is worth noting that these calculations for the $2^+$ resonance have been performed by keeping the same three-body force used to adjust the $0^+$ state to the available experimental energy. However, it has been previously shown that the three-body potential required to reproduce the known information on three-body energies might be different between different angular momentum states of a given system (see, for instance, Refs.~\cite{MRoGa04,Nguyen12,JCasal14}). Therefore, the predicted values for the energy and width of a $2^+$ resonance are somewhat arbitrary. This calls for new experimental insight to better constrain three-body models for $^{16}$Be.

\section{Summary and conclusions}
\label{sec:conclusions}
A new method to identify and characterize resonances in a discrete basis has been presented. The formalism is based on the sensitivity of resonant states to changes in the potential operator. Resonances are identified as non-stationary states, which are eigenstates of the resonance operator $\widehat{M}=\widehat{H}^{-1/2}\widehat{V}\widehat{H}^{-1/2}$ corresponding to large negative eigenvalues. The properties of the resonance, i.e., its energy and width, are obtained from the time evolution of the state by introducing a resonance quality parameter. 

The method has been tested for the unbound three-body system $^{16}$Be, of timely interest in the context of two-nucleon emitters. The hyperspherical formalism has been used to describe $^{14}\text{Be}+n+n$ continuum states, using the analytical THO basis within the PS approach. The binary potentials employed in the calculations were the GPT $n-n$ interaction and a phenomenological core-$n$ potential adjusted to the available information of the $^{15}$Be system. To reproduce the $2n$ separation energy of 1.35 MeV, an additional three-body force was included. The lowest $0^+$ eigenvalue of the resonance operator has been identified as the ground-state of $^{16}$Be. The converged wave function presents a strong dineutron configuration, confirming the findings from previous theoretical works that favor the direct two-neutron emission. From the time evolution, a width of $\Gamma(0^+)=0.16$ MeV is extracted, which is consistent with previous $R$-matrix calculations of the actual three-body scattering states. This narrow $0^+$ ground state exhibits a small deviation with respect to an ideal Breit-Wigner resonance. Following the same procedure, a $2^+$ excited state in $^{16}$Be is also predicted. Using the same three-body Hamiltonian, the resonance appears at 2.42 MeV and shows a width of 0.40 MeV. New experimental data are required to confirm the existence of this $2^+$ resonance and better constrain the three-body model for $^{16}$Be. In particular, the large discrepancy between the experimental and theoretical widths needs to be addressed. The formalism has been also applied to the $1^-$ continuum states, where no resonance is predicted. 

It should be stressed that this formalism produces a normalizable wavefunction associated to the resonance. This will allow  to use reaction-theory calculations to obtain quantitative values for the cross sections to populate the resonance from different reaction channels. 
The present formalism can be easily extended to study the resonance properties of three-body systems comprising several charged particles, for which the solution of the actual three-body scattering problem poses a challenge. Calculations for $^6\text{Be}(^4\text{He}+p+p)$ and $^{11}\text{O}(^9\text{C}+p+p)$, the unbound mirror partners of the halo nuclei $^6$He and $^{11}$Li, are in progress.

\begin{acknowledgments}
The authors would like to thank F.~M. Marqués for useful discussions and valuable insight. This work has been partially supported by the Spanish Ministerio de Ciencia, Innovaci\'on y Universidades  and the European Regional Development Fund (FEDER) under Projects  No.~FIS2017-88410-P, FPA2016-77689-C2-1-R and FIS2014-51941-P, and by the European Union's Horizon 2020 research and innovation program under grant agreement No.~654002.
\end{acknowledgments}

\bibliography{bibfile}

\end{document}